\documentclass[a4paper,11pt]{article}
\usepackage{graphicx}
\usepackage{dcolumn}
\newcommand{\beq}{\begin{equation}}
\newcommand{\eeq}{\end{equation}}

\def\mdPi{monad$_{\rm \Pi}$ }
\title{Lorentz transformation in Quantum Mechanics}
\author{M. Baldo}
\date{ }
\begin{document}
\maketitle
\vspace {-1 cm}
\begin{center}
\textit{Istituto Nazionale di Fisica Nucleare}
\par \textit{via Santa Sofia 64, 95123 Catania, Italy} 
\end{center}
\vskip 0.5 cm
\noindent Abstract.
\par 
The compatibility of special relativity and Quantum Mechanics has been questioned by several authors. The origin of this tension can be traced back mainly to the introduction of the measurement processes and the corresponding wave function reduction, which play a crucial role in Quantum Mechanics. We approach this problem with the help of a recent proposal for a model of Quantum Mechanics, where the measurement is explicitly described as a specific stochastic process. This implements
ordinary Quantum Mechanics, where measurement and reduction are treated as phenomenological events of unknown origin without any physical justification. To state clearly the question in general, we first discuss and establish the effect of a Lorentz transformation on a generic wave function in space-time. Alongside the analysis we formulate the relativistic version of the model. We then consider few thought experiments in order to analyze to what extent Quantum Mechanics
follows relativistic invariance and find the specific critical points where non invariance possibly occurs.
The analysis can shade light to the interpretation of the existing experimental observations.   
\section{Introduction.\label{sec:int}}
Quantum Mechanics (QM) and special relativity are the basis of the theoretical description of all physical phenomena, with the exception of the gravitational ones. For a coherent theory of natural phenomena it is essential that they are compatible. 
The well established relativistic Quantum Field Theory (QFT) \cite{Weinberg} is indeed a coherent formulation of QM in the relativistic framework.
However, when the wave functions are introduced to represent physical systems, evidences of possible incompatibility clearly arise, in particular when measurements are considered in space-time. 
Since the laws of ordinary Quantum Mechanics (QM) cannot be questioned, because they are in perfect agreement with experiments,
it is of great interest to clarify to what extent the laws of special relativity is compatible with them. This is surely not a new subject, but there are points that are missing in the literature, which are worth to be considered and discussed. 
The paper does not present new theory or numerical predictions or to "solve" the paradoxes,
since they are independent of any model. 
The most known example is the spin correlation measurement
between entangled particles, the so called EPR paradox \cite{EPR,Bohm}. The quantum mechanical calculation of the correlation perfectly agrees with the experimental finding, but it is performed assuming instantaneous reduction of the wave function, which amounts to an instantaneous "propagation" of the correlation. Of course a correlation is not a signal, since the result of the spin measurement is stochastic and it cannot be controlled to send a message. However it looks difficult to accommodate this instantaneous correlation within a relativistic formalism. In fact, as noticed in ref. \cite{Gao}, the order in which one of the particle affects the state of the other depends on the reference frame if Lorentz transformation (LT) are used, which hints a violation of the "cause - effect" temporal order. As it is well known \cite{Friis,Friis2} the correlation itself is not relativistic invariant since it depends on the reference frame, but the violation of Bell's inequality and of the local realism is still valid. This and some thought experiments will be discussed in section \ref{sec:measure}.
\par 
In this paper we approach the problem with the help of the model of ref.
\cite{IJQF} for the completion of QM. The model is devised to describe the physical processes that are kept outside the standard formalism, in particular the strong measurements. The basic assumption
is the introduction for each wave function of a nonstandard component, in the sense of nonstandard analysis,
and of a nonstandard space-time lattice which is the support of the wave function. The lattice has infinitesimal steps, where the time step is of higher order than the space step.
Under a set of additional assumptions, that are  coherent with the basic one, it turns out that a particular Ito stochastic process is automatically present in the standard sector. This stochastic dynamics has to be added to the standard linear evolution ruled by Schr\"odinger equation. The corresponding stochastic differential equations is then obtained under the
only condition of the conservation of the norm.  This equation is able to produce both the wave function reduction process
and the corresponding Born' s rule for the probability, which in ordinary QM are without explanations.  
The introduction of the lattice and the corresponding stochastic dynamics is the basic element of the model, since no standard external perturbation can produce the reduction of a wave function \cite{Pas}.\par
Possible predictions of the model, that could give evidence for the presence of the nonstandard lattice, are briefly discussed
in refs. \cite{IJQF,IJQF2}.\par   
The paper is organized as follows. In section \ref{sec:unit} we sketch the relativistic extension of the model. In secs. \ref{sec:lattice},\ref{sec:wp} we establish in general the LT of a wave function on the basis of QFT. In sec. \ref{sec:measure} we study few cases where measurements in space-time play an essential role and we discuss to what extent and in which sense QM and relativity are compatible.

\section{Remarks on the Lorentz transformations.\label{sec:unit}}
The Lorentz transformations (LT) leave invariant the space-time distance between two points (events).
If we call $ (t,\mathbf{x}) $ and $ (t\,',\mathbf{x}\,') $ the components of the four-vector 
joining the two points in two different reference frames, then
\beq
\mathbf{x}\,^2 \,-\, t\,^2 \,=\, \mathbf{x}\,'\,^2 \,-\, {t\,'}\,^2
\label{eq:inverse}\eeq
\noindent with $ \mathbf{x} = (x,y,z) $ and $ \mathbf{x}\,' = (x\,',y\,',z\,') $
The equality necessarily assumes the same unit of length in the two reference frames, otherwise
the equality would be ill defined. In other words, the unit of length has been calibrated between the two
references, as in all other references. This can be achieved simply by endowing each reference frame with 
an identical standard unit of length. Due to the invariance of the velocity of light, the same applies to the 
unit of time. \par 
The LT of the vector for a boost along the $ x $ axis can be written
\beq
\begin{array}{cc}
x\,' &\,=\, \frac{ x\,-\,\beta t }{\sqrt{1\,-\,\beta\,^2}} \\

t\,' &\,=\, \frac{ t\,-\,\beta x }{\sqrt{1\,-\,\beta\,^2}} 
\end{array} 
\label{eq:expli}\eeq
\noindent with $ y\,' = y$, $z\,' = z$, where $\beta = v/c$, being $v$ the relative velocity of the two frames. For simplicity we took $ c = 1 $. If one takes two events that occur at the same time in the unprimed frame, $ t = 0 $ and 
$ x = x\,' \sqrt{1\,-\,\beta\,^2} $. This is the contraction of length. Notice that correspondingly 
$ t\,' = \beta x / \sqrt{1\,-\,\beta\,^2 } $, which means that the two events are not simultaneous in the primed
reference frame. However, if the space distance is constant in time in the primed frame, the value of $ t\,'$ is irrelevant.
This is the case when the two space-time points are the extreme of a stick at rest in the primed frame. This means that the 
space contraction is a real phenomenon. As an example we can take the collision between two protons in the LHC accelerator,
where their size can be strongly contracted in the laboratory frame, i.e. they behave, just before the collision, almost
as two colliding small discs.\par
\section{Lorentz transformations and the nonstandard sector. \label{sec:lattice}}
At the classical (non-quantal) level it would be difficult to question LT, therefore we will assume that they are valid in the standard sector, i.e. for the standard part of the space-time coordinates. On the other hand we cannot use LT for the nonstandard part, because this would mix the order of the infinitesimal steps in the space-time lattice.
In fact LT involves both space and time intervals, and the infinitesimal of higher order for the time step is vanishing
small with respect to the infinitesimal for the space step. As can be seen from the LT equations (\ref{eq:expli}) the transformed time step would become of the same order of the space step, and the model would be contradictory. 
 This requires a more refined interpretation of the meaning of the space-time lattice. First of all   
it has to be stressed that the LT of Eq. (\ref{eq:expli}) include an implicit assumption, often omitted. As explicitly mentioned in the original paper by A. Einstein \cite{orig}, it is supposed that identical copies of a basic ruler and a basic clock are provided to each reference frame, that are used to measure the space-time coordinates. The values of the space-time coordinates of an event are in fact obtained by taking the ratio of the intervals along the axis from the origin and the basic
space-time units. This is a essential assumption, otherwise it would have no meaning to relate the coordinates of two different reference frames.
This is equivalent to consider these basic elements not affected by the LT. To illustrate this point, one can refer to the example of the previous section about the contraction of length, where the measure of a stick length in the moving frame
is obtained from the interval between the stick extremities taken at the same time. The interval has to be compared with
the unit of length, which of course is not taken as the LT of the one of the rest frame, but it is just the one belonging to the moving frame. 
\par 
It is natural to identify the space-time infinitesimal steps as the basic ruler and basic "clock tick", which are therefore independent of the reference frame. In this way in different reference frames an event has different space-time standard coordinates, in agreement with LT, but the same infinitesimal steps for the nonstandard lattice. This interpretation does not violate Lorentz invariance, since it is in accordance with the interpretation of the standard LT. This identification is equivalent to consider the infinitesimal steps as the minimal space-time intervals one can consider in the space-time manifold.
As explained above, this interpretation of the space-time steps is the only one that preserves the model, any other choice
would lead to inconsistent results. 
 Being infinitesimal, they are of course not directly measurable, but they can affect specific physical phenomena, as in the non-relativistic case. Because the finite part of the nonstandard straight line (or any curve) is the disjoint union of all the \mdPi of the real axis, each coordinate can be seen as the juxtaposition of an unlimited number of these basic elements along the space-time axis. All the juxtapositions which end in a given \mdPi give the same standard value for the coordinate.
\section{Lorentz transformations of a free wave packet.\label{sec:wp}} 
To simplify the exposition let us assume that the two frames of the LT have the origin of the space-time coordinates that
coincide at time $t = t' = 0$, so that the four vectors which appear in Eq. (\ref{eq:expli}) are just the coordinates of a single
point. Then the free wave packet for a spin zero and neutral particle will be a function only of the four vector $ x = (t,\mathbf{x})$. As explained in the Appendix it can be represented as a superposition of plane waves
\beq
\phi(t,\mathbf{x}) \equiv \phi(x) = \int \frac{ d^3 k }{k_0} w(\mathbf{k}) \exp(\imath k x ) 
\label{eq:wp}\eeq  
\noindent where $ k = (k_0,\mathbf{k}) $ is a generic four vector in momentum space ($\hbar$ = 1),
$k_0 = \sqrt{\mathbf{k}\,^2 + m\,^2}$, being $m$ the mass of the particle. Since the particle is a scalar the corresponding
wave function must be a scalar and its LT can be written \cite{Weinberg}
\beq
\Lambda \phi(x) \,=\, \phi(\Lambda x) \,=\, \phi(x\,')
\label{eq:LT}\eeq
\noindent which means that in the new reference frame the wave function is the same original function in
the new coordinates. It has to be stressed that the time $t\,'$ is different for each 
$\mathbf{x}$. In QFT for a field extended in a finite space-time region this non-uniqueness of the time under LT is an unavoidable consequence. It means simply that the field at a given time in a reference frame is related by a LT to the field taken at different times in another reference frame. However, if  $\phi(x)$ is meant to  represent the wave function of a particle in a given reference frame, the effect of a LT to another reference frame poses a straightforward question. In fact,
in QM the wave function at a given time is assumed to represent the quantum state of the system at that definite time.
The non-uniqueness of the time implies the impossibility of connecting by a LT a definite quantum state in two different reference frames. Indeed, in general no correspondence can be established between the quantum state represented by the wave function  $\phi(t,\mathbf{x})$ and the quantum state represented by a wave function $\phi(t\,',\mathbf{x\,'})$ at a given $t\,'$. A correspondence can be present only between the whole evolution of the wave function as observed in the two reference frames. The main reason is that the wave packet is an extended object in 3-D space and evolves in both reference frames, which marks a difference with the case of a rigid body at rest in one of the two reference frames. A further obvious difference is the impossibility to "observe" a wave function like a physical object.  
This is one of the main conclusions of the paper, which is not explicitly present in the literature.
It  is a straightforward consequence of how a LT acts on a wave function and the interpretation of the wave function in QM as representing a quantum state. Stated in a different way, if one performs a LT on a wave function from a reference frame, in the new space-time coordinates the wave functions at different times appear, and no quantum state can be defined. 
Despite that, the range of time involved in the set of transformed wave functions can be easily estimated.
Let us consider a wave function in a given reference frame $\phi(t,x)$ at a given time $t$ which has a space extension of the order $\Delta x$. A LT to a reference frame moving with velocity $\beta$ will introduce a set of wave functions spanning a
time interval of the order $\Delta t \,=\, \beta \gamma \Delta x $, where $\gamma = 1/\sqrt{1 - \beta^2}$.
\par 
In some limiting case the correspondence between the states in the two reference frames can be established, at least approximately. Suppose that the wave packet is "almost" two dimensional, i.e. in one direction the size of the wave packet can be neglected. Then in the reference frame where the x-axis is along this special direction all the points of the  wave packet have the same x-coordinate and a boost along x axis will bring to a wave packet with only one time coordinate, i.e.  a legitimate quantum state. Similarly, it can happen that in definite physical conditions the overall size of the wave packet is small enough that the spread of the time values under the LT of Eq. (\ref{eq:LT}) can be neglected.
This is the case of e.g. the semi-classical approximation.  \par 
The final conclusion is that the overall time evolution of a system is the only quantum object that can be transferred by LT from one reference frame to another. In this sense it is a covariant quantity.

\section{Measurement. \label{sec:measure}}
A strong measurement is characterized by the stochastic reduction of the wave function. The dynamics of this process is described by the stochastic differential equation derived in ref. \cite{IJQF}. It is a consequence of the presence of the nonstandard space-time lattice.
 Since each reference frame is equipped with the same nonstandard lattice, the reduction follows the same laws as in the non-relativistic case in each reference frame where a measurement takes place, as demanded by the relativity principle, provided the physical conditions are the ordinary ones, in particular the measurement apparatus is at rest. This is just the requirement that the physical laws are the same in all inertial reference frames. \par 
Let us consider the elementary process of detection of a particle by an apparatus which is at rest in the considered reference frame, e.g. a measurement of position of a particle described by a wave packet with a finite space extension. The measurement will produce the reduction of the wave function, e.g. the localization of the wave function in a restricted space region.
As seen by another reference frame, which is moving with respect to that reference frame, it will be described by performing a
LT on the space-time coordinates. Then, as discussed in the previous section \ref{sec:wp}, we have to apply the transformation to the whole evolution of the wave function. Therefore the LT will involve the wave function both before and after the reduction time in the original reference frame. As a consequence the reduction process as seen by the moving frame will not be anymore instantaneous, but it could be described in general as a set of wave functions that include a gradual transformation of the incoming wave packet into the reduced form. 
However the new description 
of the process as a set wave functions cannot be related to a quantum state evolution, since no quantum state can be indeed identified. On the other way around, if we consider the wave function as seen by the moving frame at a given time, 
by LT from the  fixed reference frame it will receive contributions from both reduced and non-reduced wave functions.   
Of course this change of description occurs for any other process that can happen to the original wave function. 
The reference frame where the experimental apparatus is at rest plays therefore a special role. In this sense the reduction process is not completely relativistic invariant \cite{Gao}. This conclusion does not depend on the particular model used for the reduction process, provided it is instantaneous in the special reference frame, but it is just a consequence of the spatial extension of the wave function.     
\par
Non locality is an intrinsic feature of QM, as it was stressed by Einstein in one of the discussion sessions
of the 1927 Solvay Conference. The Proceedings are reported in e.g. Bacciagaluppi and Valentini book \cite{Bag}. 
 Einstein presented a thought but feasible experiment, that we briefly summarize here.
He schematized the experiment with a drawing similar to the one in Fig. 1. 
\begin{center}
\begin{figure}
\includegraphics[width=1.0\textwidth]{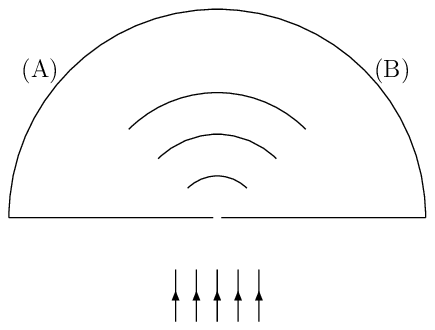} 
\vskip -12 cm
\caption{Schematic representation of the thought experiment illustrated by A. Einstein, as reported in ref. \cite{Bag}.}
\end{figure}
\end{center}
\vskip -1 cm
An electron beam is sent toward a screen with a hole small enough
that the outgoing electrons are spread in all direction in almost isotropic way. The hole then acts as a source of electrons emitted in all directions in the outgoing hemisphere. A distant photographic screen that covers the hemisphere detects the electrons at one of the possible positions, indicated by (A). The beam is supposed to have a low enough intensity that only one electron at a time is present in the region of the experiment. According to QM each outgoing electron is described by a wave packet with an angular spread covering the hemisphere, while at the screen the electron is localized at (A). Before detection the wave packet is spread around all the portions of the screen, so potentially it could be detected at any other point (B). The detection at (A) delete this possibility instantaneously, so something "has happened" in (B). This suggests some sort of non locality, but of course this cannot be tested, since the detection in (A) or in (B) are mutually exclusive. In any case this way of thinking is directly related to the finite size of the wave packet and to the integrity of the wave function.
N. Bohr was present in the audience and admitted that it had not understood what Einstein wanted to say. 
The non locality apparent in this thought experiment is the consequence of the wave function reduction under measurement and the wave function spacial extension.   
However, it has to be noticed that the reduction is the result \cite{IJQF} of the stochastic process among the components of the wave function, which takes place in Hilbert space, i.e. among the corresponding amplitudes. In the particular case of Fig. 1 the detector is measuring the position of the particle and the components are just the localized states  
all around the screen where e.g. a scintillator is positioned. Therefore the "disappearance" of the wave function in (B)
is a consequence of the stochastic dynamics that occurs in the abstract space and not in ordinary space. 
For this reason a LT has not to be applied to the different points of the wave function which is instantaneously "vanishing" in (B) and reduced around (A). In fact, a process of "disappearance" of the wave function in some region of space would not be instantaneous in a generic reference frame, but at different time in different points of the wave function. In other words the reduction of the wave function produced by the position measurement must not be viewed as a collapse of the shape and extension of the wave function. This feature is a peculiarity of measurement and reduction processes. The modification of the
wave function by standard interaction processes can be followed in space time coordinates, since they occur explicitly 
in ordinary space, where the Schrodinger or the corresponding relativistic equations apply. In any case the LT on standard processes, as noticed above, can be performed only on the whole evolution of the system.         
Of course, the LT from one reference frame to another amounts to apply a LT to the space-time coordinates of the position of detection if one can neglect the size of the "spot" at the screen. Furthermore a possible observation of the wave function reduction in space-time is ruled out, since any observation of the evolution in a single measurement would remove the detection at the screen.
It has to be remembered that in the case of non-zero spin, the particle state is modified by a LT, because the spin or the photon polarization are not relativistic invariant,
and the output of the measurement depends on the reference frame.   \par
The expected result that the possible approximate space-time position of a particle detection follows LT under change of reference frame has the 
straightforward consequence in the case of several measurements on a given system. The sequence of the measurements can depend on the reference frame. In particular if two particles are detected simultaneously in a reference frame at different positions,
they are not simultaneous in any other reference frame and their sequence can be different. This is obvious since the space-time interval between the two detections is space-like. As an example one can consider the case of an hypothetical decay of a particle at rest into two equal mass particles in opposite directions. If in that reference frame two detectors are placed at equal distances at opposite sites they will detect the two particles simultaneously, but the sequence of the two detections in another reference frame will depend on the direction of its motion. In the non-relativistic QM framework the wave function $\Psi$ of the two emitted particles can be written as the product of the wave function $\Phi$ of the center of mass motion and of the relative motion $\phi$
\beq
\Psi(\mathbf{r_1},\mathbf{r_2}) \,=\, \Phi(\mathbf{R}) \phi(\mathbf{r})                                   
\label{eq:prod}\eeq 
\noindent where $\mathbf{R} = (\mathbf{r_1} + \mathbf{r_2})/2$ and $\mathbf{r} = \mathbf{r_1} - \mathbf{r_2}$ and we neglect spin for simplicity. Let us neglect the momentum distribution associated with the center of mass wave function, which can be in fact minimal under proper physical conditions. Its inclusion would not affect the picture and the conclusions.
The total momentum and angular momentum of the two emitted particle are then vanishing, therefore the relative wave function is in the s-wave state. Expanding in the relative momentum $\mathbf{k}$ one can write (forgetting normalization)
\beq
\phi(\mathbf{r}) \,=\, \int k^2 dk j_0(kr) g(k) e^{2\imath E_k t} \,=\, \int k^2 dk d\Omega_k e^{\imath(\mathbf{k}\mathbf{r_1} - \mathbf{k}\mathbf{r_2})} g(k) e^{2\imath E_k t }
\label{eq:swave}\eeq
\noindent where $j_0$ is the spherical Bessel function of order zero and $g(k)$ takes into account the energy width of the decaying particle. Eq. (\ref{eq:swave}) describes a spherical wave packet expanding outwards.
If the detectors are sensitive to momentum, the detection of e.g. particle 1 will select its momentum, leaving the undetected particle $2$ in the state
\beq
\phi_2(\mathbf{r_2}) \,=\, \exp(-\imath \mathbf{k}\mathbf{r_2}) g(k) e^{\imath E_k t }
\label{eq:det1}\eeq
\noindent where $\mathbf{k}$ is the selected momentum. Therefore particle $2$ will be in a plane wave of opposite momentum 
$-\mathbf{k}$. Then the quantum description follows closely the classical one, in particular the effect of LT.\par 
A different scenario appears if the detectors and the decaying particle are not aligned. Let us consider the situation depicted schematically in Fig. 2, i.e. the detectors point towards the decaying particle at 90° with respect to each other. 
\begin{center}
\begin{figure}[h]
\includegraphics[width=1.0\textwidth]{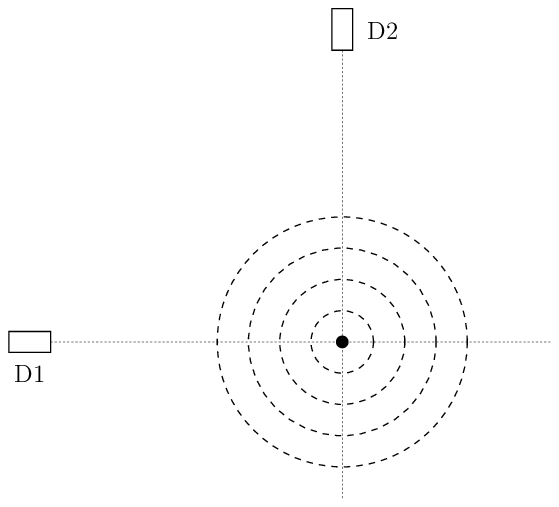}
\vskip -12 cm
\caption{Schematic representation of the thought experiment discussed in the text. The boxes indicate two detectors.
The dashed circles represent pictorially the spherical wave packet of the relative motion of two emitted particles.}
\end{figure}
\end{center}
\vskip -1 cm
If the distance of detector 2 from the decaying particle is slightly higher than detector 2, particle 1 will be detected before particle 2, forcing particle 2 to be into the state of opposite momentum, as described above. In this case particle 2 will miss the detector 2, which therefore will remain silent. This arrangement can be considered as the two-particle version of the one in Fig. 1. In this case it is possible to check the "disappearance" of the wave function in a place away from the one of the detection. 
\par  
A LT could reverse the order of the possible detections, in which case detector 2 will fire, while detector 1 will be silent.
In more details, let us consider first the reference frame where the detectors are at rest and arrange the detectors at equal distances from the source. Then the two particles will arrive at the corresponding detectors simultaneously. Let us first switch off the detectors, so we can consider the two arrivals without the occurrence of detection. If we now perform a LT along the x-axis, i.e we observe the system from a moving frame, the two arrivals will be not anymore simultaneous,
but they will be observed with a time delay $\Delta t = \gamma\beta \Delta x $, where
$ \Delta x$ is the distance along the x-axis of the two detectors in the original standing frame. Then, according to the sign of the velocity $\beta$ between the two frames, the arrivals will be in different temporal order. Switching on the detectors, this means that only one of the detectors will be operating.   
\par 
Along these lines one can consider the EPR \cite{EPR} experiments, which had numerous realizations and improvements in order to finally circumvent all possible pitfalls and loopholes. An extensive list of references on the experiments can be found in ref. \cite{loop}. Of course the main interest on these experiments is the observed violation of Bell's inequality, which excludes the possibility of a local realism in QM. The origin of the correlations involved in the Bell' s inequality is essentially threefold, the reduction of the two particle wave function under measurement, the extension of the wave function and its integrity. The measurement selects one two-particles spin (or polarization) state between the two that appears in the entangled superposition.
The effect of LT on the values of the spin correlations where considered extensively in ref. \cite{Friis,Friis2}. Here we focus 
on the space-time correlations of the measurements by the two opposite detectors, not considered in \cite{Friis,Friis2}.
The QM calculation of the correlations assumes an instantaneous reduction and it is in perfect agreement with observations,
for all the detector distances considered in the experiments, in particular when the detectors are at equal distance from the source. This gives support to the use of the wave function in the QM formalism, which then acquires some degree of realism. This result do not depend on possible model for the reduction process, it must be only compatible with the instantaneity assumption, like the model of ref. \cite{IJQF}. Let me stress that instantaneity property is not a trivial 
requirement, since, to the author knowledge, hardly other models are able to 'explain' it.
\par 
In the Bohm's version of EPR experiments, two spin-1/2 particles are emitted
from a decaying particle in a singlet state in opposite directions. If we neglect the fluctuations of the center of mass, the state can be written
\beq
\frac{1}{\sqrt{2}} \big[\phi(t,\mathbf{r},\uparrow)\phi(t,-\mathbf{r},\downarrow) \,-\,\phi(t,\mathbf{r},\downarrow)\phi(t,-\mathbf{r},\uparrow)\big]
\label{eq:sing}\eeq  
\noindent where the decaying particle is at rest at $\mathbf{r} = 0$ and the two $\phi$'s are wave packets well localized around the two classical (non-quantum) positions at time $t$. The localization is assumed to be strong enough that in the following we can neglect the effects of the wave packet quantum extension. Two detectors sensitive to the spin direction are positioned at opposite sides along the direction
of the particles. Then the arrivals of the two particles at the corresponding detectors will be simultaneous. As it is well known the spins of the two particles are correlated. The origin of this entanglement is the particular superposition of Eq. (\ref{eq:sing}), which implies that if one of the particle is detected with spin up the other must have spin down, and vice versa. As already mentioned this correlation is assumed to be instantaneous. \par 
Let us now perform a LT, as schematically represented in Fig. 3, to a reference frame moving with velocity $V$ along 
the x-axis.
\begin{center}
\begin{figure}[h]
\includegraphics[width=1.0\textwidth]{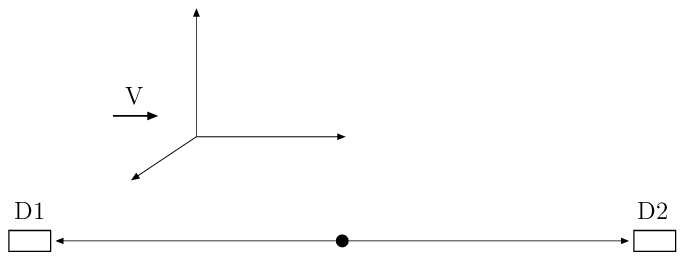}
\vskip -12 cm
\caption{Schematic representation of the thought experiment discussed in the text. The boxes indicate two detectors. A particle at the position indicated by a black dot decays into two identical particles in opposite directions.
A LT is performed to a reference frame with a velocity $V$ along the x-axis. As observed in this frame the arrivals of the two particles are not anymore simultaneous. As soon as the detector D2 is reached by a particle and the corresponding spin is measured, the spin of the other particle is fixed before it can reach D1.  }
\end{figure}
\end{center}
\vskip -1 cm
\par 
As already discussed above, the arrivals of the two particles at the detectors are not any more simultaneous. One particle will be detected at D2, and its spin selected, before the other particle will reach D1, with a time delay $\Delta t = \gamma\beta \Delta x $, where its spin will be also measured. However the selection of the spin of this second particle occurs before it can reach D1.
Following ref. \cite{Gao}, if we now consider another reference frame moving in the opposite direction, i.e. with velocity $-V$, the roles of D1 and D2 will be reversed. Each observer in the two moving frames will assume that the spin selection in the corresponding detector that is first firing is stochastic, with 50\% probability for each spin direction. However this assumption is illusory since the two spins are correlated. These considerations suggest that the measurements in the two detectors has to be pictured as a single measurement on the two particle systems, whose overall output is indeed stochastic. 
This consideration does not solve the problem that arises if one takes the correlation between the two particles as a cause-effect relation, since in any situation the physical cause must be prior to the effect \cite{Gao}. The dependence on the observer is a straightforward consequence of the space-like character of the space-time interval between the two detections. It is a well known statement in relativity that no cause-effect relation can exist between events that are at space-like distance, since this would involve a physical action with a velocity higher than the speed of light. The process of wave function reduction has to be considered  a special one, which does not follow the laws valid for physical objects, in particular special relativity.  This can be considered the major point of tension between QM and relativity, which however is not surprising since the wave function cannot be considered a physical object in strict sense. \par 
Another point to be stressed is about the application of Born's rule by an observer that is moving with respect the measuring devices. As discussed above, the wave function before the measurement simply does not exist in the moving frame. The observer must then refer the experiment to the proper time of the reference frame of the detectors, where one can apply the Born's rule. The corresponding probabilities are the only ones that can be calculated and by construction they are relativistic invariant.       
\section{Summary and conclusions}
In the realistic interpretations of Quantum Mechanics one introduces the wave function as a fundamental tool at the basis of its formalism and of its practical applications. Historically the wave function was devised in order to explain the observations of diffraction and interference of elementary particles, in particular electrons. Since then it became the universal mathematical object in the formalism to be used to interpret and design all the experiments where Quantum Mechanics  
is involved. In the applications a special role is played by the measurement processes, where the so called wave function reduction occurs. The reduction is not part of the mathematical formalism but introduced as a separate postulate without an explicit description. Recently a model was proposed \cite{IJQF} that extends the standard Quantum Mechanics to incorporate in the formalism the reduction process. In order to devise the relativistic version of the model, we first establish the effect of a LT on a generic wave function. 
On this basis we discuss the LT of the reduction process and we consider several thought experiments, to enlight possible incompatibility between relativity and Quantum Mechanics. One of the main feature of the LT on a wave function is the impossibility of establishing a direct relation between the observations of a quantum state in two different reference frames.
Only the overall evolution of a wave function in two reference frames can be connected by a LT. The reason for this peculiarity is just the unavoidable changes under LT of the time interval between two non coincident space-time events.
As a consequence, in the case of a generic detection, the reduction time, which is vanishing in the detector reference frame, appears finite under a LT. The length of this time depends on the size of the wave function undergoing the reduction.\par 
Analyzing some thought experiments some apparent paradoxical situations seem to occur. In particular, in a two-particle systems, the arrangement of two detectors can be such that only one particle is detected, but which one of the two depends on the reference frame. In other words, the only detector that is actually firing depends on the reference frame. Despite this can appear paradoxical, it does not violate any basic principle. We finally considered the EPR experiment, and it is pointed out that the corresponding reduction process is not relativistic invariant, and therefore a special role is played by the reference frame where the experimental apparatus is at rest and where the probabilities, according to the Born' s rule, have to be calculated. Despite that, if a model, like the one proposed in ref. \cite{IJQF}, is able to explain the reduction of a general wave function in a measuring process, there is no logical or factual paradox. These conclusions specify to what extent Quantum Mechanics and relativity are compatible.        
\section{Appendix.\label{sec:LT} }
Let us consider neutral spinless field $\psi(t,\mathbf{x})$ which obeys the Klein-Gordon equation. The field can be expanded 
\beq
\psi(t,\mathbf{x}) \,=\, \int  \frac{d^3 p}{(2\pi)^\frac{3}{2}\sqrt{2 k_0}} \big( e^{\imath k x} a (\mathbf{k}) \,+\,
e^{-\imath k x} a^\dagger (\mathbf{k}) \big)
\label{eq:exp}\eeq
\noindent where the operators $a^\dagger (\mathbf{k})$ and $a (\mathbf{k})$ are the creation and annihilation of the quanta of the field, i.e. of the corresponding particle momentum state $| \mathbf{k} > $. At $ t = 0 $ the application of the field operator to the vacuum state $ | 0 > $ gives
\beq
\psi(0,\mathbf{x}) | 0 > \,=\, \int  \frac{d^3 p}{(2\pi)^\frac{3}{2}\sqrt{2 k_0}} e^{-\imath \mathbf{k} \mathbf{x}} | \mathbf{k} > 
\label{eq:vac}\eeq
\noindent Since the states $ | \mathbf{k} > $ are eigenstates of the free hamiltonian with energy $k_0$, the free evolution of the state (\ref{eq:vac}) can e written
\beq
\psi(t,\mathbf{x}) | 0 > \,=\, \int \frac{d^3 p}{(2\pi)^\frac{3}{2}\sqrt{2 k_0}} e^{-\imath k x} | \mathbf{k} >
\label{eq:free}\eeq
\noindent The expansion of Eq. (\ref{eq:free}) can be inverted
\beq
\frac{1}{\sqrt{2 k_0}} e^{-\imath k_0 t} | \mathbf{k} > \,=\, \frac{1}{(2\pi)^\frac{3}{2}} \int d^3 x
e^{-\imath \mathbf{k} \mathbf{x}} \psi(t,\mathbf{x}) | 0 >
\label{eq:mom}\eeq
\noindent For a generic quantum state $ | \Psi > $ of the free motion one gets
\beq
\frac{1}{(2\pi)^\frac{3}{2}} \int d^3 x
e^{-\imath \mathbf{k} \mathbf{x} } < \Psi | \psi(t,\mathbf{x}) | 0 > \,=\, < \Psi | \tilde{\mathbf{k}}, t > 
\label{eq:scal}\eeq
\noindent where
\beq
| \tilde{\mathbf{k}}, t > \,=\, \frac{1}{\sqrt{2 k_0}} e^{-\imath k_0 t} | \mathbf{k} > 
\label{eq:momt}\eeq
\noindent The left and right amplitudes can be identified with the wave functions (wave packet) in the coordinates and momentum representations respectively, related by the usual Fourier transform. Inverting Eq. (\ref{eq:scal}) one gets the equivalent of Eq. (\ref{eq:wp}). The orthogonality between the momentum states has the relativistic form
\beq
 < \tilde{\mathbf{k'}}, 0 | \tilde{\mathbf{k}}, 0 > \,=\, \frac{1}{2 k_0} \delta( \mathbf{k} - \mathbf{k'} )
\label{eq:ort}\eeq
\noindent As it is well known (Weinberg) the field operator under the action of a Lorentz transformation $ \Lambda $ changes as
\beq
\Lambda \psi(x) \Lambda^{-1} \,=\, \psi(\Lambda x) 
\label{eq:fieldLT}\eeq
\noindent Applying this relation to (\ref{eq:scal}) one gets Eq. (\ref{eq:LT}).

\end{document}